\documentclass[11pt]{article}
\include{epsf}

\def\appendix#1{\addtocounter{section}{1}\setcounter{equation}{0}
\renewcommand{\thesection}{\Alph{section}}
\section*{
\thesection\protect\indent \parbox[t]{11.715cm} {#1}}
\addcontentsline{toc}{section}{Appendix\thesection\ \ \ #1} }

\def\be{\begin{equation}}
\def\ee{\end{equation}}
\def\bea{\begin{eqnarray}}
\def\eea{\end{eqnarray}}

\begin{document}

\author{Ralf Lehnert\\
{\it CENTRA, \'Area Departamental de F\'{\i}sica,
Universidade do Algarve}\\
{\it 8000-117 Faro, Portugal}\\
rlehnert@ualg.pt}

\title{Fundamental physics and Lorentz violation}

\date{October 2003}
\maketitle
\begin{abstract}
The violation of Lorentz symmetry
can arise in a variety of approaches to fundamental physics.
For the description of the associated low-energy effects,
a dynamical framework known as the Standard-Model Extension
has been developed.
This talk gives a brief review of the topic
focusing on Lorentz violation through varying couplings.
\end{abstract}

\section{Introduction}
\setcounter{equation}{0}

On the one hand,
the Standard Model (SM) of particle physics
is extremely successful phenomenologically.
On the other hand,
this conference is called
{\it What Comes Beyond the Standard Model}
because it is generally believed
that the SM is really the low-energy limit
of a more fundamental theory
incorporating quantum gravity.
Experimental research in this field
faces various challenges.
They include
the expected Planck suppression of quantum-gravity signatures
and the absence of a realistic underlying framework.

A promising approach for progress
in quantum-gravity phenomenology
is the identification of relations
that satisfy three principal criteria:
they must hold exactly in known physics,
they are expected to be violated in candidate fundamental theories,
and they must be testable with ultra-high precision.
Spacetime symmetries satisfy all of these requirements.
Lorentz and CPT invariance are key features
of currently accepted fundamental physics laws,
and they are amenable to Planck-sensitivity tests.
Moreover,
Lorentz and CPT breakdown
has been suggested in a variety of approaches to fundamental physics.
We mention
low-energy emergent Lorentz symmetry \cite{np83},
strings \cite{kps},
spacetime foam \cite{ell98},
nontrivial spacetime topology \cite{klink},
loop quantum gravity \cite{amu},
noncommutative geometry \cite{chklo},
and varying couplings \cite{klp03}.
The latter of these mechanism will be discussed
in more detail in this talk.

At presently attainable energies,
Lorentz and CPT violating effects
are described
by a general extension of the SM.
The idea is to include into the SM Lagrangian
Lorentz and CPT breaking operators
of unrestricted dimensionality
only constrained by coordinate independence \cite{sme}.
This Standard-Model Extension (SME)
has provided the basis
for many investigations
placing bounds on Lorentz and CPT violation.
For the best constraints in the matter and photon sectors,
see Ref.\ \cite{bear} and Refs. \cite{cfj,km},
respectively.
Note
that certain Planck-suppressed SME operators
for Lorentz and CPT breaking
provide alternative explanations
for the baryon asymmetry in our universe \cite{bckp}
and the observed neutrino oscillations \cite{neu}.

\section{Lorentz violation through varying couplings}
\setcounter{equation}{0}

Early speculations in the subject of varying couplings
go back to Dirac's numerology \cite{lnh}.
Subsequent theoretical investigations have shown
that time-dependent couplings arise naturally
in many candidate fundamental theories \cite{theo}.
Recent observational claims
of a varying fine-structure parameter $\alpha$ \cite{webb}
have led to a renewed interest in the subject \cite{jpuzan}.

Varying couplings are associated
with spacetime-symmetry violations.
For instance,
invariance under temporal and/or spatial translations
is in general lost.
Since translations are closely interwoven
with the other spacetime transformations
in the Poincar\'e group,
one anticipates
that Lorentz symmetry
might be affected as well.
This
is best illustrated by an example.
Consider the Lagrangian ${\cal L}$
of a complex scalar $\Phi$,
and suppose a spacetime-dependent parameter $\xi(x)$
is coupled to the kinetic term:
${\cal L}\supset\xi\partial_{\mu}\Phi\partial^{\mu}\Phi^*$.
An integration by parts
yields
${\cal L}\supset-\Phi(\partial_{\mu}\xi)\partial^{\mu}\Phi^*$.
If,
for instance,
$\xi$ varies smoothly on cosmological scales,
$(\partial_\mu\xi)=k_\mu$
is essentially constant locally.
The Lagrangian
then contains a nondynamical fixed 4-vector $k_\mu$
selecting a preferred direction
in the local inertial frame
violating Lorentz symmetry.

The above example can be generalized
to other situations.
For instance,
non-scalar fields can play a role,
and Lorentz violation can arise
through coefficients like $k_\mu$
in the equations of motion
or in subsidiary conditions.
Note
that the Lorentz breaking
is independent
of the chosen reference frame:
if $k_\mu\neq 0$
in a particular set of local inertial coordinates,
$k_\mu$ is nontrivial in any coordinate system.
In the next section,
we show
that varying couplings can arise
through scalar fields acquiring
expectations values in a cosmological context.
Note,
however,
that the above argument for Lorentz violation
is independent of the mechanism
driving the variation of the coupling.

\section{Four-dimensional supergravity cosmology}
\setcounter{equation}{0}

Consider a Lagrangian ${\cal L}_{\rm sg}$ with two real scalars $A$ and $B$
and a vector $F^{\mu\nu}$:
\begin{eqnarray}
\frac{4{\cal L}_{\rm sg}}{\sqrt{g}}
&=&
\frac{{\partial_\mu A\partial^\mu A
+ \partial_\mu B\partial^\mu B}}{B^2}
- 2 R
- M F_{\mu\nu} F^{\mu\nu}
- N F_{\mu\nu} \tilde{F}^{\mu\nu} \; ,
\nonumber\\
M &=& \frac{B (A^2 + B^2 + 1)}{(1+A^2 + B^2)^2 - 4 A^2}\; ,\quad
N = \frac{A (A^2 + B^2 - 1)}{(1+A^2 + B^2)^2 - 4 A^2}\; ,\qquad
\label{sugra}
\end{eqnarray}
where $g^{\mu\nu}$ represents the graviton
and $g=-\det (g_{\mu\nu})$, as usual.
We have denoted the Ricci scalar by $R$,
the dual tensor is
$\tilde{F}^{\mu\nu}=\varepsilon^{\mu\nu\rho\sigma}F_{\rho\sigma}/2$,
and the gravitational coupling has been set to one.
Then,
the Lagrangian (\ref{sugra})
fits into the framework of the pure $N=4$ supergravity
in four spacetime dimensions.

To investigate Lagrangian (\ref{sugra}) in a cosmological context,
we assume a flat Friedmann-Robertson-Walker universe
and model galaxies and other fermionic matter
by including the energy-momentum tensor $T^{\mu\nu}$ of dust,
as usual.\footnote{The dust can be accommodated
into the supergravity framework,
which also contains fermions uncoupled from the scalars.}
In such a situation,
the equations of motion can be integrated analytically \cite{klp03}
yielding a nontrivial dependence of $A$ and $B$
(and thus $M$ and $N$)
on the comoving time $t$.
Comparison with the usual electrodynamics Lagrangian
in the presence of a $\theta$ angle shows
$\alpha \equiv 1/4\pi M(t)$ and $\theta \equiv 4\pi^2 N(t)$,
so that the fine-structure parameter
and the $\theta$ angle
acquire related time dependences in our supergravity cosmology.

If mass-type terms
${\cal L}_m=-\sqrt{g}(m_A A^2 + m_B B^2)/2$
for the scalars
are included into Lagrangian (\ref{sugra}),
our simple model
can match
the observed
late-time
acceleration of the cosmological expansion \cite{klp03}.
Note also
that the scalars themselves
obey Lorentz-violating dispersion relations \cite{klp03,rl03}.

\section*{Acknowledgments}
I wish to thank C.\ Froggatt,
N.\ Manko\v{c}-Bor\v{s}tnik,
and H.B.\ Nielsen
for organizing a stimulating meeting.
This work was supported in part
by the the Centro Multidisciplinar de Astrof\'{\i}sica (CENTRA),
the Funda\c{c}\~ao para a Ci\^encia e a Tecnologia (FCT),
and an ESF conference grant.

\end{document}